\documentclass{aastex631}

\shorttitle{Close KBO Binaries}
\shortauthors{Porter et al.}

\begin{document}

\title{Detection of Close Kuiper Belt Binaries with HST WFC3}

\author[0000-0003-0333-6055]{Simon B. Porter}
\affil{Southwest Research Institute, 1050 Walnut St, Suite 300, Boulder, CO 80302, USA}

\author[0000-0001-8821-5927]{Susan D. Benecchi}
\affiliation{Planetary Science Institute, Tucson, AZ, USA}

\author[0000-0002-3323-9304]{Anne J. Verbiscer}
\affiliation{University of Virginia, Charlottesville, VA, USA}

\author[0000-0002-8296-6540]{W. M. Grundy}
\affiliation{Lowell Observatory, Flagstaff, AZ, USA}

\author[0000-0002-6013-9384]{Keith S. Noll}
\affiliation{NASA Goddard Spaceflight Center, Greenbelt, MD, USA}

\author[0000-0002-6722-0994]{Alex H. Parker}
\affiliation{SETI Institute, Mountain View, CA, USA }

\begin{abstract}
    Binaries in the Kuiper Belt are common.
    Here we present our analysis of the Solar System Origins Legacy Survey (SSOLS) 
    to show that using a PSF-fitting method can roughly
    double the number of binaries identified in that dataset.
    Out of 198 Kuiper Belt objects (KBOs) observed by SSOLS, we find 23 to be visually separated
    binaries, while a further 19 are blended-PSF binaries detectable with the method
    we present here.
    This is an overall binary fraction of 21\% for the SSOLS dataset of cold classical KBOs.
    In addition, we tested our fitting methods on synthetic data,
    and while we were able to show it to be very effective at detecting certain blended-PSF 
    binary KBOs, fainter or closer binary KBOs may easily be missed,
    suggesting that the close binary KBO fraction could be even higher.
    These results strongly support the idea that most (if not all) KBOs were formed through
    the Streaming Instability process, and as a consequence, most KBOs
    were formed as near-equal mass binaries.
\end{abstract}

    \received{5/31/2023}

    \revised{4/2/2024}

    \accepted{4/14/2024}


    \submitjournal{Planetary Science Journal}


\section{Introduction} \label{sec:intro}

The Solar System is a highly evolved planetary system that betrays few clues of its origin, with the key exception of the Kuiper Belt.
The present day Kuiper Belt consists almost exclusively of objects that formed outside of the giant planets and were either pushed outward
by giant planet migration (the Neptune-resonant and scattered populations) or are still in the original orbits where they formed around the Sun
\citep[the classical KBOs;][]{2020Sci...367.6620M}.
KBOs, and especially the cold classical KBOs (CCKBOs) which have the least perturbed heliocentric orbits,
have suffered very little bombardment since their formation relative to asteroids interior of Jupiter \citep{2021JGRE..12606961M},
and no real thermal processing like comets \citep{2020Sci...367.3705G}.
The best constraints on the surfaces of KBOs come from the New Horizons mission, which flew past both the Pluto system and the CCKBO Arrokoth.
Pluto itself is a highly-processed world with extensive surface-atmosphere interactions \citep{2015Sci...350.1815S}.
Pluto’s large satellite Charon and two of four small satellites (Nix and Hydra) were imaged at sufficient resolution to show a small amount of cratering \citep{2017Icar..287..187R},
but since all the satellites were likely formed after a giant impact \citep{2011AJ....141...35C},
that says little about Kuiper Belt history as a whole.
Arrokoth, a typical CCKBO \citep{2018AJ....156...20P}, was much more constraining.
It is constructed of two distinct lenticular lobes, connected on their equators, forming a contact binary \citep{2020Sci...367.3999S}.
There is a single large crater on the smaller lobe, and all other craters are significantly smaller \citep{2020Sci...367.3999S}.
This clean contact binary shape is unlike any object previously encountered by a spacecraft, but provides tantalizing clues to the formation processes of the solar system 
\citep{2020Sci...367.6620M}.

A leading theory of the formation of the Outer Solar System is the Streaming Instability 
\citep[SI;][]{2019NatAs...3..808N}.
This postulates that solid bodies in the circumsolar disk initially formed when small clumps of solid material caused localized increases in gas drag,
allowing the small clumps to very rapidly grow to full sized KBOs within tens of years \citep[and citations therein]{2019NatAs...3..808N}.
An important consequence of this rapid growth (when compared to more traditional slow accretion models) is that angular momentum grows rapidly too.
SI thus tends to produce binary systems (and occasionally triple and higher systems), with most of the angular momentum in the mutual orbits \citep{2017NatAs...1E..88F}.
In addition, because the source angular momentum is primarily from Keplerian shear, binaries produced by SI should be generally prograde \citep{2019NatAs...3..808N}.
And because SI requires the gas disk to happen, some fraction of SI-created binaries should have evolved inwards under gas drag until they became contact binaries
\citep[as may have happened for Arrokoth;][]{2020Sci...367.6620M},
while others evolved under mutual and solar tides to become tight circular binaries \citep{2012Icar..220..947P}.
SI thus predicts a Kuiper Belt filled with binaries of varying mutual separations, primarily prograde orbits, as well as a significant fraction of contact binaries.

The Kuiper Belt actually observed does indeed contain many binaries, which are primarily on prograde orbits \citep{2019Icar..334...62G}.
Photometric lightcurve studies appear to show that at least 15\% of CCKBOs are contact binaries \citep{2019AJ....157..228T},
though this number could be an underestimate due the difficulty of detecting contact binaries from lightcurves \citep{2021Icar..35614098S}.
Similar studies of the (closer and brighter) 3:2 Neptune resonant ``Plutinos'' found a contact binary fraction of up to 50\%
\citep{2018AJ....155..248T}.
Arrokoth is particularly instructive, as its dual-lenticular contact binary shape is best explained by formation as a
binary by SI and slow evolution to contact under gas drag \citep{2020Sci...367.6620M}.
The case for SI to be the leading formation mechanism for the Kuiper Belt, and the Outer Solar System generally, is thus very strong.
However, a key prediction of SI is there should be a very large number of very tight ($<$3000 km separation) binary KBOs.
The overall fraction of KBOs represented by these tight binaries is unknown, but must be significant to hide a large fraction of SI-produced binaries below the 
angular resolution of HST and ground-based adaptive optics systems.
Measuring the separations of very tight binary KBOs (and especially CCKBOs) thus provides critical constraints not only on the SI,
but on the processes that have sculpted the orbits of binary KBOs since then, including solar perturbations \citep{2012Icar..220..947P}
and gas drag \citep{2020Sci...367.6620M}.

The Wide Field Camera 3 (WFC3) on the Hubble Space Telescope (HST) offers the best available tool to detect binary KBOs, 
as it is able to observe very faint objects with very high angular precision. 
Here we show we can detect and characterize binaries that are so close that they not clearly separated in HST/WFC3 images. 
This is aided by the fact that the WFC3 point spread function (PSF) is on average consistent and very well characterized, 
allowing for a precise measurement of a binary KBO with a separation of at least 0.5 pixels, 
as and bright as V=24.
We apply this method to 198 CCKBOs observed with HST program GO-15648 (the Solar System Origins Legacy Survey, SSOLS),
and show that it can roughly double the yield of binaries relative to those which can be easily detected with visual inspection.
We also use this model to create 1000 synthetic KBO cases implanted in real HST images, and show the limits of this
detection method.

\begin{figure}
\plotone{\detokenize{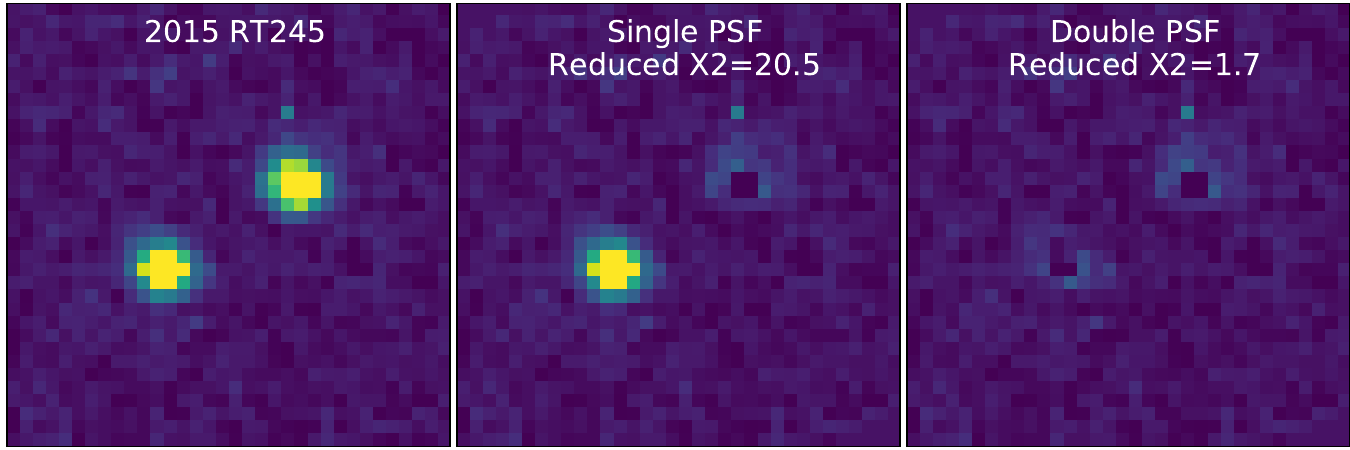}}\\
\plotone{\detokenize{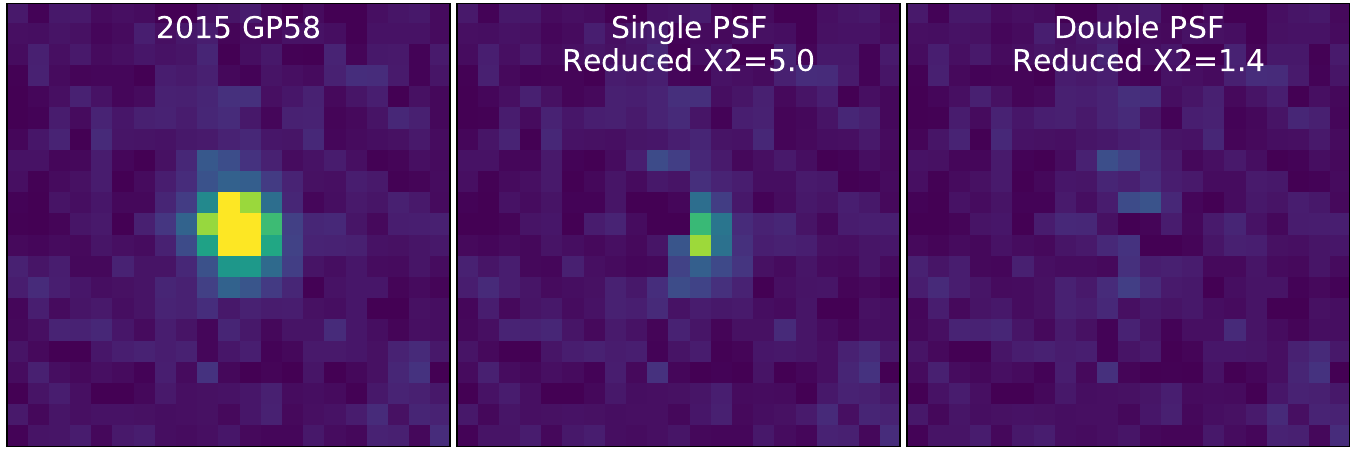}}\\
\plotone{\detokenize{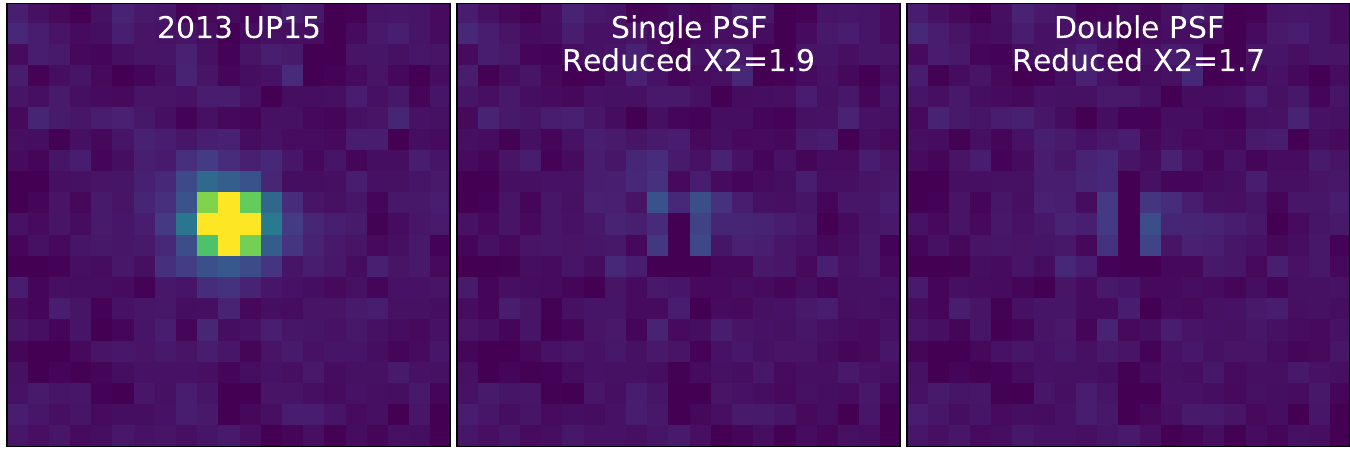}}
\caption{
    Examples of (top) a well-separated binary KBO, (middle) a blended binary KBO, and (bottom) a KBO that appears to be single.
    The top separated binary case accounted for 11.6\% of the KBOs in the SSOLS dataset.
    For the middle case,
    the single-PSF solution only subtracts the primary, leaving behind a significant residual which is not there for the
    double-PSF solution, strongly indicating that this KBO is a blended binary.
    The residuals for the bottom case are almost identical for the single-PSF and double-PSF solutions, indicating that it
    is not binary to the resolution of HST/WFC3.
    All of these images are the F606W stacks for these objects; we performed the same analysis for the F814W stacks
    and enforced consistency between them for objects that we identified as blended binaries.
    \label{fig:sepbin}
}
\end{figure}

\section{Target Selection and HST Imaging}

The Cycle 26 Treasury Program `The Solar System Origins Legacy Survey' (SSOLS) observed 198 CCKBOs at optical wavelengths.
This sample drew on three successful ground-based discovery surveys, 
the Deep Ecliptic Survey \citep[DES;][]{2005AJ....129.1117E},
the Canada France Ecliptic Plane Survey \citep[CFEPS;][]{2011AJ....142..131P}, 
and the Outer Solar System Origins Survey \citep[OSSOS;][]{2018ApJS..236...18B}.
The DES was the first deep systematic survey to probe the population of the Kuiper Belt,
and discovered 320 KBOs and Centaurs, including 43 of the SSOLS targets \citep{2005AJ....129.1117E}.
CFEPS probed 321 degree$^2$ of the sky, discovering and characterizing 169 KBOs, discovering 4 of the SSOLS targets \citep{2011AJ....142..131P}.
OSSOS was a followup of CFEPS, discovering over 800 KBOs and Centaurs, including 151 of the SSOLS targets \citep{2018ApJS..236...18B}.
The SSOLS targets were chosen to be cold classical KBOs that did not appear to be binary in the ground-based imagery,
with the exception of 2005 EO$_{304}$, the widest binary in the SSOLS target list.
This sample set of CCKBOs was designed to provide identifiable CCKBO binaries at a level that
allows discovery bias to be quantified and removed, to better understand the binarity and luminosity function of the intrinsic CCKBO population.
All of the targets (listed in Table \ref{tab:list}) had well-determined orbits with at least three oppositions of data,
to allow confidence in both their dynamical classification and that would be observable by the Hubble Space Telescope (HST)
with a small subarray.

The 198 KBOs were successfully observed with HST program GO-15648, ranging in approximate apparent magnitude from V=21 to V=25.
Observations of an additional 14 objects were attempted, but failed due to HST not correctly locking on to its guide star,
leaving the observations unusable.
The target CCKBOs were between 35.98 and 53.43 astronomical units from Earth at the time they were imaged,
corresponding to a spatial resolution in the WFC3 images of 1044 to 1550 km/pixel.
Each target KBO was observed with a single orbit, with four images in the F606W filter (roughly V-band), 
and four images in the F814W filter (roughly R-band), all using the UVIS2-C512C-SUB subframe. 
A four-point box dither was used for each filter.
These filters were chosen to provide the most amount of color information while still maximizing the 
signal-to-noise ratio (SNR) for binary detection.
F606W and F814W have also been used for previous HST observations of CCKBOs, particularly in support of the New Horizons
mission \citep{2019Icar..334...22B}, allowing these observations to place both the color distribution of the binaries in context
of the broader cold classical Kuiper Belt and the well-characterized New Horizons Distant KBO subset.
Here we only present the results of relative astrometry and binary identification, while a following paper
(in prep) will present the color results.
An additional paper (in prep) will use these binaries to present a model for the size frequency distribution 
of binaries in the CCKBOs.

All of the data presented in this paper were obtained from the Mikulski Archive for Space Telescopes (MAST) at the Space Telescope Science Institute. 
The specific observations analyzed can be accessed via \dataset[doi:10.17909/6xbd-kd20]{https://doi.org/10.17909/6xbd-kd20}. 

\startlongtable
\begin{deluxetable*}{rl|ccccc}
    \tablecaption{
        The SSOLS target CCKBOs, their F606W magnitudes, their combined relative probabilities of being binary 
        (Equation \ref{eq:rpt}), the best-fit separation of the double-PSF case, the best-fit flux ratio of the
        the double-PSF case, and our judgement if the object is a separated binary, blended binary, or single.
        \label{tab:list}
    }
    \tablehead{
        & & \colhead{F606W Mag} & \colhead{Bin. Prob.}
        & \colhead{Pix. Sep.} & \colhead{Flux Ratio} & \colhead{Binary?}
    }
    \startdata
(612087) & 1999 CU$_{153}$ & 23.4 & 0.891 & 1.0 & 2.0:1 &  \\
(439858) & 1999 ON$_{4}$ & 23.6 & 0.425 & 1.1 & 2.5:1 &  \\
(612157) & 2000 FG$_{8}$ & 24.1 & 0.634 & 0.9 & 2.8:1 &  \\
(455206) & 2001 FE$_{193}$ & 23.0 & 0.082 & 1.5 & 5.7:1 &  \\
(612213) & 2001 FK$_{185}$ & 23.3 & 0.022 & 2.1 & 15.2:1 &  \\
(385362) & 2002 PT$_{170}$ & 23.3 & 0.014 & 2.7 & 9.0:1 &  \\
(469509) & 2003 HC$_{57}$ & 22.2 & 1.000 & 1.5 & 1.9:1 & Blended \\
(612549) & 2003 HG$_{57}$ & 22.8 & 1.000 & 7.6 & 1.5:1 & Separated \\
(363401) & 2003 LB$_{7}$ & 22.6 & 0.007 & 1.9 & 100.0:1 &  \\
(612619) & 2003 SN$_{317}$ & 22.9 & 0.998 & 0.8 & 2.0:1 & Blended \\
& 2003 UK$_{293}$ & 23.4 & 0.186 & 1.2 & 10.2:1 &  \\
& 2003 UV$_{291}$ & 24.0 & 0.006 & 1.8 & 100.0:1 &  \\
(612733) & 2003 YU$_{179}$ & 23.0 & 1.000 & 1.1 & 1.7:1 & Blended \\
(444018) & 2004 EU$_{95}$ & 23.0 & 0.001 & 6.2 & 100.0:1 &  \\
& 2004 HD$_{79}$ & 22.5 & 1.000 & 2.1 & 1.4:1 & Separated \\
& 2004 HE$_{79}$ & 23.5 & 0.717 & 0.9 & 2.5:1 &  \\
(469610) & 2004 HF$_{79}$ & 22.9 & 1.000 & 3.2 & 1.4:1 & Separated \\
& 2004 HG$_{79}$ & 23.1 & 0.629 & 1.0 & 2.6:1 &  \\
(444025) & 2004 HJ$_{79}$ & 23.3 & 0.958 & 1.3 & 2.6:1 & Blended \\
& 2004 HK$_{79}$ & 23.4 & 1.000 & 3.1 & 1.1:1 & Separated \\
& 2004 KE$_{19}$ & 22.7 & 1.000 & 3.3 & 2.2:1 & Separated \\
& 2004 KF$_{19}$ & 22.7 & 0.043 & 4.1 & 54.2:1 &  \\
& 2004 KG$_{19}$ & 23.3 & 0.841 & 0.8 & 2.4:1 &  \\
& 2004 MU$_{8}$ & 23.0 & 1.000 & 4.3 & 1.3:1 & Separated \\
& 2004 PT$_{117}$ & 23.1 & 0.984 & 1.0 & 2.3:1 & Blended \\
& 2004 PU$_{117}$ & 23.2 & 0.089 & 1.1 & 6.6:1 &  \\
& 2004 PV$_{117}$ & 22.8 & 1.000 & 5.8 & 5.6:1 & Separated \\
& 2004 PW$_{117}$ & 23.3 & 1.000 & 17.5 & 2.0:1 & Separated \\
& 2004 PX$_{117}$ & 23.5 & 1.000 & 4.4 & 1.5:1 & Separated \\
(609222) & 2004 VB$_{131}$ & 22.9 & 0.903 & 0.9 & 2.4:1 &  \\
(609221) & 2004 VC$_{131}$ & 22.0 & 0.294 & 1.3 & 2.2:1 &  \\
& 2004 VD$_{131}$ & 23.0 & 0.974 & 0.8 & 1.5:1 & Blended \\
& 2005 BW$_{49}$ & 22.8 & 1.000 & 2.4 & 1.7:1 & Separated \\
& 2005 CE$_{81}$ & 23.8 & 1.000 & 6.1 & 1.1:1 & Separated \\
(525461) & 2005 EN$_{302}$ & 23.7 & 0.001 & 5.4 & 100.0:1 &  \\
(525462) & 2005 EO$_{304}$ & 22.6 & 1.000 & 43.1 & 4.2:1 & Separated \\
& 2006 CH$_{69}$ & 23.2 & 1.000 & 23.0 & 1.7:1 & Separated \\
& 2006 JV$_{58}$ & 23.3 & 1.000 & 3.6 & 1.7:1 & Separated \\
& 2006 QA$_{181}$ & 23.5 & 0.010 & 3.0 & 9.9:1 &  \\
(587670) & 2006 QE$_{181}$ & 23.8 & 0.658 & 0.6 & 1.7:1 &  \\
& 2006 QF$_{181}$ & 23.4 & 0.999 & 0.8 & 1.6:1 & Blended \\
(523615) & 2006 UO$_{321}$ & 23.7 & 0.088 & 1.4 & 9.4:1 &  \\
& 2006 WF$_{206}$ & 23.1 & 1.000 & 1.3 & 1.3:1 & Blended \\
& 2007 CQ$_{79}$ & 23.4 & 0.884 & 1.1 & 3.6:1 &  \\
& 2007 CS$_{79}$ & 23.3 & 0.466 & 0.9 & 2.8:1 &  \\
& 2007 DS$_{101}$ & 22.8 & 1.000 & 2.1 & 1.3:1 & Separated \\
& 2013 EM$_{149}$ & 23.0 & 1.000 & 1.3 & 2.0:1 & Blended \\
(500840) & 2013 GA$_{138}$ & 23.9 & 0.483 & 0.7 & 1.4:1 &  \\
& 2013 GB$_{138}$ & 23.6 & 0.459 & 0.9 & 4.0:1 &  \\
& 2013 GC$_{138}$ & 23.6 & 0.219 & 1.9 & 12.4:1 &  \\
& 2013 GD$_{138}$ & 24.3 & 0.024 & 2.1 & 5.5:1 &  \\
& 2013 GE$_{138}$ & 24.3 & 0.053 & 1.5 & 5.9:1 &  \\
& 2013 GF$_{138}$ & 23.5 & 0.894 & 1.1 & 3.7:1 &  \\
(500835) & 2013 GN$_{137}$ & 22.9 & 0.997 & 1.2 & 2.5:1 & Blended \\
& 2013 GP$_{137}$ & 23.8 & 0.005 & 1.7 & 100.0:1 &  \\
(500836) & 2013 GQ$_{137}$ & 23.5 & 0.076 & 1.5 & 5.9:1 &  \\
& 2013 GR$_{137}$ & 24.2 & 0.284 & 0.6 & 3.2:1 &  \\
& 2013 GS$_{137}$ & 23.9 & 0.317 & 0.9 & 3.1:1 &  \\
(500837) & 2013 GT$_{137}$ & 24.0 & 0.369 & 1.0 & 6.2:1 &  \\
& 2013 GU$_{137}$ & 23.7 & 0.668 & 0.7 & 1.5:1 &  \\
(500838) & 2013 GV$_{137}$ & 23.3 & 0.920 & 0.7 & 2.0:1 &  \\
(500839) & 2013 GW$_{137}$ & 23.8 & 0.070 & 1.6 & 4.5:1 &  \\
& 2013 GX$_{137}$ & 23.5 & 0.586 & 0.9 & 2.6:1 &  \\
& 2013 GY$_{137}$ & 23.7 & 0.130 & 0.8 & 2.5:1 &  \\
(500856) & 2013 HT$_{156}$ & 24.7 & 0.009 & 4.7 & 6.4:1 &  \\
& 2013 SC$_{101}$ & 24.1 & 0.036 & 0.9 & 100.0:1 &  \\
& 2013 SD$_{101}$ & 23.8 & 0.039 & 1.7 & 6.3:1 &  \\
& 2013 SE$_{101}$ & 24.1 & 0.340 & 1.0 & 2.8:1 &  \\
& 2013 SF$_{101}$ & 23.9 & 0.046 & 1.5 & 3.8:1 &  \\
& 2013 SG$_{101}$ & 24.5 & 0.042 & 2.4 & 11.4:1 &  \\
& 2013 SJ$_{100}$ & 24.5 & 0.628 & 0.9 & 6.8:1 &  \\
& 2013 SL$_{100}$ & 23.8 & 0.854 & 0.9 & 2.2:1 &  \\
& 2013 SO$_{100}$ & 24.1 & 0.073 & 2.1 & 12.6:1 &  \\
(505446) & 2013 SP$_{99}$ & 23.4 & 0.578 & 0.8 & 2.3:1 &  \\
(505447) & 2013 SQ$_{99}$ & 23.4 & 1.000 & 3.4 & 1.5:1 & Separated \\
& 2013 SV$_{100}$ & 23.9 & 0.766 & 0.8 & 2.3:1 &  \\
& 2013 SX$_{100}$ & 24.2 & 0.028 & 2.3 & 4.2:1 &  \\
& 2013 UB$_{18}$ & 24.2 & 0.076 & 2.2 & 6.5:1 &  \\
& 2013 UC$_{18}$ & 24.5 & 0.040 & 1.8 & 3.3:1 &  \\
& 2013 UD$_{18}$ & 23.7 & 0.005 & 1.8 & 100.0:1 &  \\
& 2013 UG$_{18}$ & 24.3 & 0.090 & 0.5 & 5.3:1 &  \\
(505476) & 2013 UL$_{15}$ & 23.2 & 1.000 & 4.3 & 1.6:1 & Separated \\
& 2013 UL$_{17}$ & 24.4 & 0.001 & 5.9 & 100.0:1 &  \\
& 2013 UN$_{15}$ & 23.8 & 0.712 & 0.6 & 2.2:1 &  \\
& 2013 UN$_{17}$ & 24.4 & 0.116 & 1.0 & 6.5:1 &  \\
& 2013 UO$_{15}$ & 23.3 & 0.796 & 0.7 & 1.8:1 &  \\
& 2013 UP$_{15}$ & 23.6 & 0.842 & 0.6 & 2.1:1 &  \\
& 2013 UP$_{17}$ & 24.3 & 0.137 & 0.8 & 2.2:1 &  \\
& 2013 UR$_{17}$ & 24.1 & 0.917 & 0.6 & 1.9:1 &  \\
& 2013 UT$_{17}$ & 24.0 & 0.337 & 0.8 & 5.0:1 &  \\
& 2013 UW$_{16}$ & 23.6 & 0.970 & 1.1 & 1.4:1 & Blended \\
& 2013 UW$_{17}$ & 24.3 & 0.325 & 0.9 & 5.2:1 &  \\
& 2013 UY$_{16}$ & 24.1 & 0.664 & 0.8 & 2.8:1 &  \\
& 2013 UY$_{17}$ & 23.8 & 0.022 & 2.4 & 4.0:1 &  \\
& 2014 UC$_{228}$ & 24.2 & 0.394 & 1.3 & 1.6:1 &  \\
& 2014 UC$_{229}$ & 24.7 & 0.436 & 0.8 & 3.6:1 &  \\
& 2014 UC$_{230}$ & 24.5 & 0.129 & 1.7 & 8.9:1 &  \\
(511551) & 2014 UD$_{225}$ & 23.1 & 0.984 & 0.9 & 2.1:1 & Blended \\
(511552) & 2014 UE$_{225}$ & 22.7 & 0.768 & 0.5 & 1.4:1 &  \\
& 2014 UL$_{228}$ & 24.0 & 0.113 & 1.0 & 4.1:1 &  \\
& 2014 UL$_{229}$ & 24.3 & 0.003 & 6.0 & 17.3:1 &  \\
& 2014 UP$_{228}$ & 24.0 & 0.009 & 3.3 & 10.7:1 &  \\
& 2014 UY$_{228}$ & 24.2 & 0.033 & 2.0 & 4.0:1 &  \\
(523756) & 2014 WD$_{509}$ & 22.5 & 1.000 & 6.4 & 2.0:1 & Separated \\
& 2015 GA$_{57}$ & 24.2 & 0.121 & 3.5 & 30.7:1 &  \\
& 2015 GB$_{57}$ & 24.5 & 0.107 & 0.9 & 3.6:1 &  \\
& 2015 GC$_{57}$ & 24.0 & 0.480 & 1.2 & 2.7:1 &  \\
& 2015 GC$_{58}$ & 24.3 & 0.962 & 0.8 & 5.1:1 & Blended \\
& 2015 GD$_{57}$ & 23.8 & 0.022 & 0.8 & 100.0:1 &  \\
& 2015 GD$_{59}$ & 24.2 & 0.048 & 0.6 & 19.4:1 &  \\
& 2015 GE$_{57}$ & 24.5 & 0.006 & 1.7 & 100.0:1 &  \\
& 2015 GF$_{56}$ & 24.3 & 0.018 & 1.9 & 12.3:1 &  \\
& 2015 GF$_{58}$ & 23.6 & 0.034 & 2.0 & 10.1:1 &  \\
& 2015 GF$_{59}$ & 24.1 & 0.355 & 1.6 & 4.7:1 &  \\
& 2015 GG$_{57}$ & 24.6 & 0.464 & 0.9 & 2.5:1 &  \\
& 2015 GH$_{58}$ & 23.8 & 0.992 & 1.5 & 9.7:1 & Blended \\
& 2015 GJ$_{57}$ & 24.2 & 0.013 & 2.7 & 13.0:1 &  \\
& 2015 GK$_{58}$ & 24.6 & 0.383 & 1.1 & 4.9:1 &  \\
& 2015 GL$_{57}$ & 24.3 & 0.135 & 2.2 & 7.6:1 &  \\
& 2015 GL$_{58}$ & 24.3 & 0.036 & 2.1 & 4.7:1 &  \\
& 2015 GM$_{58}$ & 24.0 & 0.327 & 1.1 & 4.1:1 &  \\
& 2015 GN$_{58}$ & 24.3 & 0.154 & 2.3 & 9.4:1 &  \\
& 2015 GO$_{57}$ & 24.3 & 0.563 & 0.8 & 3.1:1 &  \\
& 2015 GO$_{58}$ & 24.6 & 0.452 & 1.2 & 2.5:1 &  \\
& 2015 GP$_{58}$ & 24.0 & 0.979 & 1.1 & 1.3:1 & Blended \\
& 2015 GR$_{56}$ & 24.1 & 0.596 & 0.8 & 3.4:1 &  \\
& 2015 GR$_{57}$ & 24.0 & 0.008 & 3.3 & 12.3:1 &  \\
& 2015 GS$_{56}$ & 24.5 & 0.853 & 1.0 & 1.9:1 &  \\
& 2015 GS$_{57}$ & 24.0 & 0.150 & 1.1 & 3.2:1 &  \\
& 2015 GT$_{58}$ & 24.4 & 0.204 & 1.4 & 3.1:1 &  \\
& 2015 GU$_{56}$ & 23.5 & 0.952 & 0.8 & 1.7:1 & Blended \\
& 2015 GU$_{57}$ & 23.9 & 0.006 & 1.6 & 100.0:1 &  \\
& 2015 GU$_{58}$ & 23.4 & 0.102 & 1.4 & 8.3:1 &  \\
& 2015 GW$_{56}$ & 24.4 & 0.037 & 1.4 & 6.5:1 &  \\
& 2015 GW$_{57}$ & 24.5 & 0.011 & 6.3 & 100.0:1 &  \\
& 2015 GX$_{56}$ & 24.3 & 0.152 & 2.0 & 7.7:1 &  \\
& 2015 GY$_{56}$ & 24.6 & 0.344 & 0.9 & 5.9:1 &  \\
& 2015 GY$_{57}$ & 24.6 & 0.102 & 1.8 & 6.8:1 &  \\
& 2015 GZ$_{56}$ & 23.5 & 0.007 & 6.6 & 100.0:1 &  \\
& 2015 GZ$_{57}$ & 24.6 & 0.018 & 0.5 & 100.0:1 &  \\
& 2015 RA$_{280}$ & 25.1 & 0.525 & 0.8 & 3.0:1 &  \\
& 2015 RB$_{280}$ & 24.3 & 1.000 & 5.6 & 1.7:1 & Separated \\
& 2015 RB$_{281}$ & 23.5 & 0.029 & 1.8 & 12.0:1 &  \\
& 2015 RC$_{281}$ & 24.1 & 0.026 & 1.9 & 7.4:1 &  \\
& 2015 RD$_{280}$ & 24.4 & 0.015 & 0.9 & 100.0:1 &  \\
& 2015 RE$_{280}$ & 24.6 & 0.025 & 2.4 & 4.5:1 &  \\
& 2015 RJ$_{277}$ & 23.1 & 1.000 & 1.2 & 2.7:1 & Blended \\
& 2015 RO$_{281}$ & 23.0 & 1.000 & 1.2 & 6.2:1 & Blended \\
& 2015 RP$_{279}$ & 24.5 & 0.040 & 1.6 & 8.4:1 &  \\
& 2015 RP$_{280}$ & 24.0 & 1.000 & 3.9 & 1.8:1 & Separated \\
& 2015 RP$_{281}$ & 25.2 & 0.439 & 0.5 & 1.4:1 &  \\
& 2015 RT$_{245}$ & 23.6 & 1.000 & 12.2 & 1.1:1 & Separated \\
& 2015 RZ$_{279}$ & 24.2 & 0.017 & 3.2 & 5.0:1 &  \\
& 2015 VA$_{169}$ & 24.3 & 0.715 & 1.1 & 3.2:1 &  \\
& 2015 VA$_{172}$ & 23.6 & 0.045 & 2.0 & 10.9:1 &  \\
& 2015 VB$_{169}$ & 24.6 & 0.001 & 6.1 & 100.0:1 &  \\
& 2015 VB$_{170}$ & 23.7 & 0.874 & 1.0 & 2.3:1 &  \\
& 2015 VB$_{171}$ & 24.6 & 0.164 & 1.1 & 4.1:1 &  \\
& 2015 VB$_{173}$ & 24.3 & 0.005 & 1.6 & 100.0:1 &  \\
& 2015 VC$_{170}$ & 23.6 & 0.286 & 0.8 & 4.5:1 &  \\
& 2015 VC$_{172}$ & 23.8 & 0.003 & 6.2 & 100.0:1 &  \\
& 2015 VD$_{169}$ & 24.1 & 0.005 & 1.7 & 100.0:1 &  \\
& 2015 VE$_{169}$ & 24.0 & 0.019 & 2.3 & 6.1:1 &  \\
& 2015 VF$_{169}$ & 24.5 & 0.036 & 1.7 & 3.4:1 &  \\
& 2015 VG$_{169}$ & 24.7 & 0.040 & 1.4 & 7.0:1 &  \\
& 2015 VH$_{169}$ & 25.2 & 0.003 & 6.9 & 58.1:1 &  \\
& 2015 VH$_{171}$ & 23.7 & 0.143 & 1.1 & 4.4:1 &  \\
& 2015 VH$_{173}$ & 23.8 & 0.024 & 1.6 & 7.6:1 &  \\
& 2015 VJ$_{170}$ & 24.7 & 0.001 & 6.0 & 100.0:1 &  \\
& 2015 VK$_{169}$ & 24.9 & 0.032 & 0.4 & 100.0:1 &  \\
& 2015 VK$_{170}$ & 24.0 & 0.023 & 2.2 & 6.8:1 &  \\
& 2015 VL$_{171}$ & 24.5 & 0.014 & 0.6 & 100.0:1 &  \\
& 2015 VM$_{173}$ & 23.2 & 1.000 & 4.0 & 1.3:1 & Separated \\
& 2015 VN$_{171}$ & 24.3 & 0.003 & 5.9 & 10.9:1 &  \\
& 2015 VN$_{172}$ & 24.3 & 0.265 & 0.9 & 2.8:1 &  \\
& 2015 VO$_{171}$ & 24.1 & 0.269 & 1.0 & 4.2:1 &  \\
& 2015 VP$_{168}$ & 24.4 & 0.547 & 0.8 & 1.8:1 &  \\
& 2015 VP$_{172}$ & 24.4 & 0.100 & 1.1 & 5.7:1 &  \\
& 2015 VP$_{173}$ & 24.3 & 0.012 & 3.7 & 9.2:1 &  \\
& 2015 VQ$_{168}$ & 24.4 & 0.002 & 4.6 & 100.0:1 &  \\
& 2015 VQ$_{169}$ & 23.9 & 0.712 & 0.8 & 2.6:1 &  \\
& 2015 VQ$_{172}$ & 23.6 & 0.121 & 1.8 & 9.1:1 &  \\
& 2015 VQ$_{173}$ & 23.7 & 0.824 & 1.0 & 2.8:1 &  \\
& 2015 VR$_{168}$ & 24.5 & 0.289 & 1.2 & 4.5:1 &  \\
& 2015 VR$_{172}$ & 23.6 & 0.315 & 1.0 & 8.6:1 &  \\
& 2015 VS$_{168}$ & 24.3 & 0.008 & 3.7 & 6.2:1 &  \\
& 2015 VS$_{172}$ & 24.3 & 0.232 & 0.8 & 2.0:1 &  \\
& 2015 VT$_{168}$ & 22.9 & 0.997 & 0.9 & 3.4:1 & Blended \\
& 2015 VU$_{168}$ & 24.1 & 0.008 & 5.5 & 17.2:1 &  \\
& 2015 VU$_{169}$ & 23.8 & 0.064 & 1.6 & 7.6:1 &  \\
& 2015 VU$_{171}$ & 24.0 & 0.006 & 1.9 & 100.0:1 &  \\
& 2015 VW$_{168}$ & 24.7 & 1.000 & 4.8 & 1.8:1 & Separated \\
& 2015 VW$_{170}$ & 24.4 & 0.005 & 9.7 & 5.0:1 &  \\
& 2015 VW$_{172}$ & 24.8 & 0.362 & 1.6 & 6.5:1 &  \\
& 2015 VX$_{169}$ & 24.2 & 0.088 & 1.5 & 5.0:1 &  \\
& 2015 VY$_{170}$ & 23.9 & 0.724 & 0.9 & 2.0:1 &  \\
& 2015 VY$_{172}$ & 24.1 & 0.069 & 1.6 & 8.1:1 &  \\
& 2015 VZ$_{169}$ & 24.6 & 0.051 & 1.6 & 4.4:1 &  \\
    \enddata
\end{deluxetable*}

\section{Image Analysis}

The first step in our image analysis was to reproject all of the images, in both filters, on to a common frame that
cancelled out the dither pattern.
Pixels affected by saturation, crosstalk, or CTE tails (as identified in the data quality part of the FLC files)
were masked out here and in subsequent analysis.
For most orbits, a 300x300 pixel initial image was sufficient to find the KBO, but a few required a 700x700 pixel window 
(i.e. the full frame with ample margin for dither).
We then manually selected the object from this stacked image and recorded the offset to allow subsequent images to be properly centered.
We next used the offsets to reproject the images again now centered on the objects with a smaller (181 pixel)
window, upsampled from the original image by a factor of two.
These second stacks were used to check if the object was a well-separated binary, as was the case for 23 of 198 KBOs.
In the case that it was a separated binary, we recorded the positions of the assumed primary and secondary objects,
and if no secondary was obvious, just the primary object.
We then produced the final reprojected images, centered on the midpoint between the objects if a secondary was manually identified,
or on the primary object if not.
These final images were separately stacked for each of the two filters, F606W and F814W.

We then fit the separate stacks for each filter with a model Point Spread Function (PSF) from the Tiny Tim package \citep{1995ASPC...77..349K}.
We independently fit both of the filters with both a single-PSF and double-PSF model, and compared the resulting $\chi^2$.
The image $\chi^2$ was estimated as the sum of the square of the difference between the model and the stack of images. 
For the single-PSF cases, we used the manual pick of the primary as the initial location for the PSF, and then optimized the
flux and x/y offset of the model PSF to minimise the image $\chi^2$.
If the object had been manually identified as a binary in the prior steps, that was used as the initial conditions for the
location of the secondary for the double-PSF cases.
If there were not an identified secondary object, we used an initial guess of the brightest pixel of the residuals
from the single-PSF fit.

If the double-PSF model showed little improvement over the single-PSF model, the two were held as equaly likely,
while if the double-PSF model was a significant improvement over the single-PSF model, we considered that to be likely a binary.
We quantified this as:

\begin{eqnarray}
    \Delta_{prob} = exp(-0.5\times(\chi^2_{double} - \chi^2_{single})) \\
    rprob_1 = \Delta_{prob}/(\Delta_{prob}+1.)
\end{eqnarray}

Where $\Delta_{prob}$ is the relative likelihood of the double-PSF solution versus the single-PSF solution,
and $rprob_1$ is the likelihood that the KBO is a binary based on the $\chi^2$ residuals.
If $\chi^2_{double}$ is almost the same as $\chi^2_{single}$, then $\Delta_{prob}$ is close to unity,
and $rprob_1$ is close to 0.5, as there is little evidence that either solution is superior.
However, if $\chi^2_{double}$ is much smaller than $\chi^2_{single}$, then $\Delta_{prob}$ is much larger 
than unity, and $rprob_1$ closer to 1.0, as there is strong evidence that the double-PSF solution is superior.

While this comparison of the $\chi^2$ values is useful for filtering out the cases that were not improved
with the double-PSF model, it is susceptible to low signal-to-noise ratio cases where the second PSF is 
fitting a coherent noise spike.
We addressed this by fitting both filters independently, and then comparing the results for consistency.
We deweighted the probability of being a binary for any case
where the location of the secondary was larger than 1/4 pixel between the filters,
or where the difference in the primary and secondary delta magnitudes was more than 1.
The quarter-pixel test was based on a few tests with both real and synthetic data.
We quantified this as:

\begin{eqnarray}
    rprob_2 = 0.25/|pixsep_{F606W} - pixsep_{F814W}| \\
    rprob_3 = 1.0/| {\Delta}mag_{F606W} - {\Delta}mag_{F814W} |\\
    rprob_{total} = rprob_1 \times rprob_2 \times rprob_3 \label{eq:rpt}
\end{eqnarray}

Where $pixsep_{F606W}$ and $pixsep_{F814W}$ are the primary-secondary separation distance in pixels for 
each of the two filters,
${\Delta}mag_{F606W}$ and ${\Delta}mag_{F814W}$ are the primary-secondary differences in magnitude for 
the two filters,
and $rprob_{total}$ is the combined estimated probability of the KBO being binary.
The constraint for ${\Delta}mag$ consistency is weaker than for $pixsep$ to allow for the secondaries to
have different F606W-F814W colors than primaries, but still useful to filter out inconsistent solutions.

\section{Calibration of Detection Efficiency}

\begin{figure}[h]
    \includegraphics{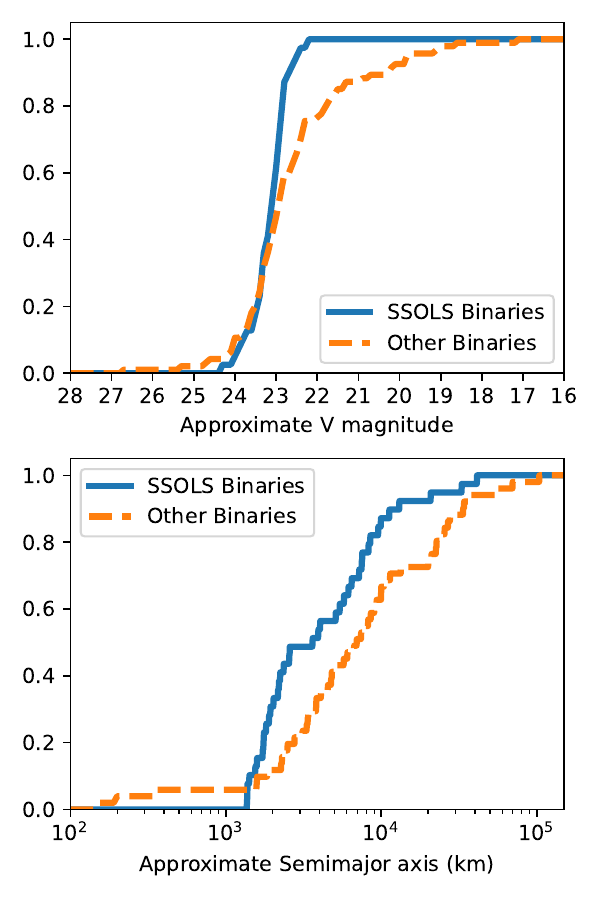}
    \caption{
        Cumulative distributions of the brightnesses and physical separations of the 
        SSOLS binaries in Table \ref{tab:list} (blue, solid) in comparison with the
        known KBO binaries (orange, dashed) listed at \url{http://www2.lowell.edu/ users/grundy/tnbs/status.html}.
        The semimajor axes of the SSOLS binaries is estimated as their apparent separation times $\sqrt{2}$.
        The SSOLS binaries are both fainter and tighter than the known binary KBOs.
        The only binary KBOs that are tighter than the SSOLS blended binaries are the 
        two detected by New Horizons \citep[2011 JY$_{31}$ and 2014 OS$_{393}$;][]{2022PSJ.....3...46W}
        and the third by a stellar occultation \citep[2014 WC$_{510}$;][]{2020PSJ.....1...48L}.
        \label{fig:cudist}
    }
\end{figure}

We sought to calibrate the detection efficiency of very close binary KBOs (and very close binary minor planets in general)
with a set of simulations using the same fitting techniques as described above on synthetic data.
Specifically, we selected four orbits from the SSOLS dataset 
(idy601, idy62h, idy64j, and idy693)
that had a corner that was clear of stars and any diffraction spikes from the target KBO.
For each of these four datasets, we then generated 250 synthetic model binary KBOs and added them to the real data in the F606W band, 
so that each synthetic binary KBO had real read noise and cosmic rays.
We varied brightnesses of the primaries of the binaries from V=21 to V=26, the separation of the binaries from 0 to 3 WFC3 pixels,
and the flux ratio between primary and secondary objects was tested at 1:1, 2:1, 5:1, and 10:1.
The model assumed 0.1 pixels 1-$\sigma$ of jitter per exposure.

\begin{figure*}[ht!]
    \plottwo{\detokenize{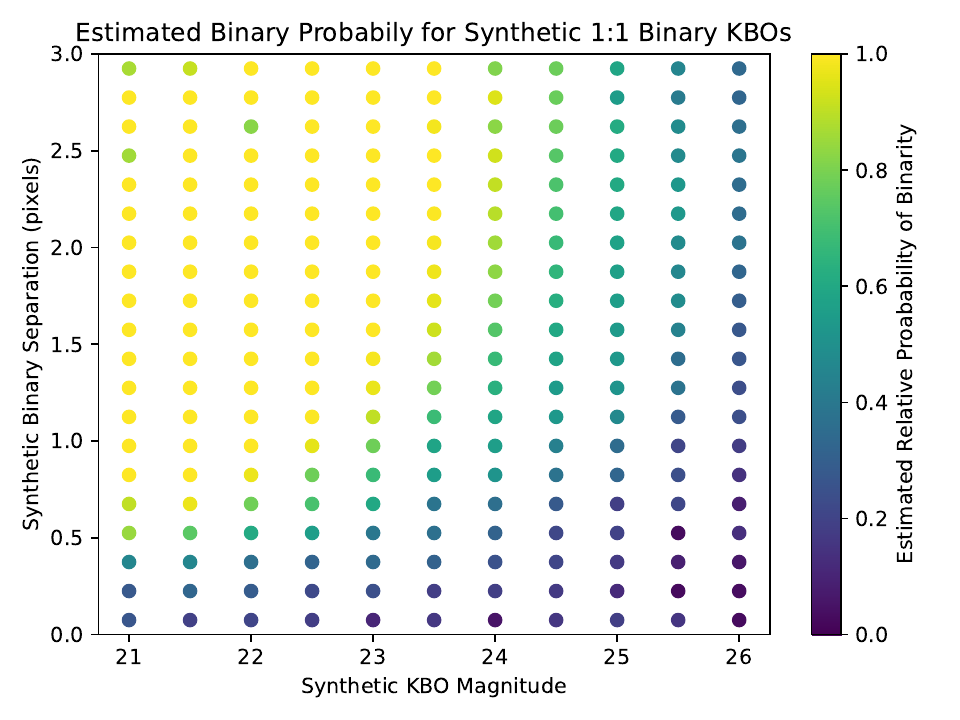}}{\detokenize{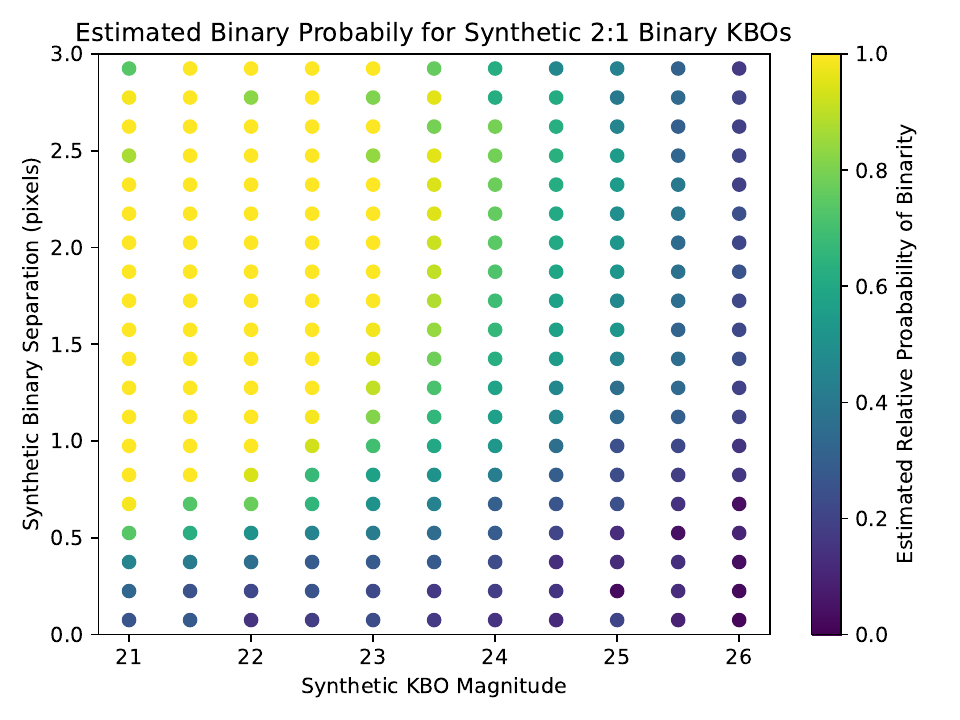}}
    \plottwo{\detokenize{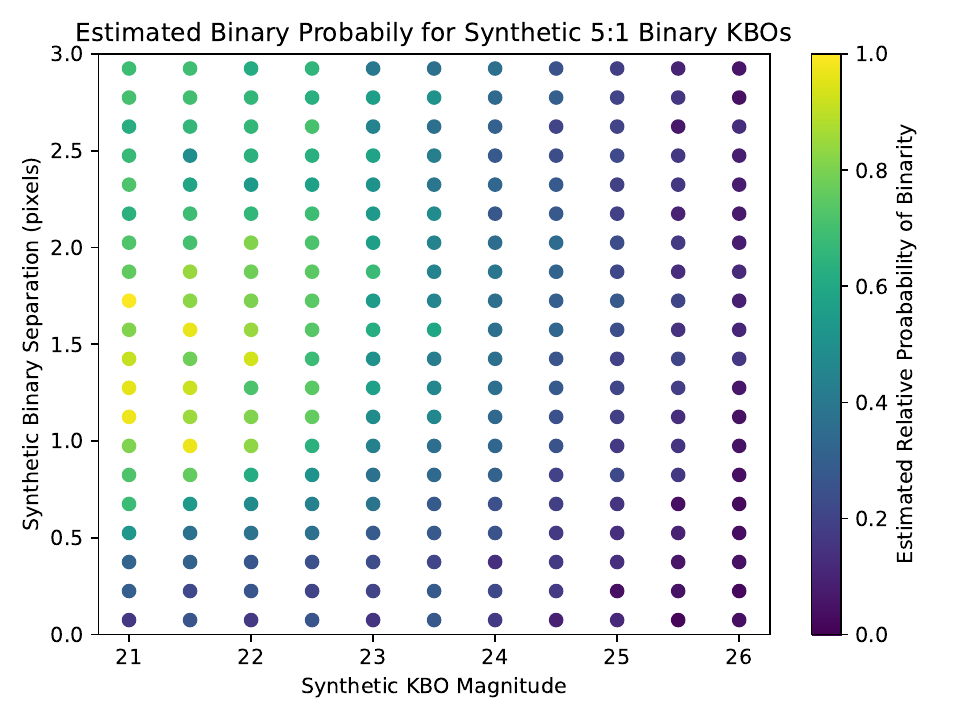}}{\detokenize{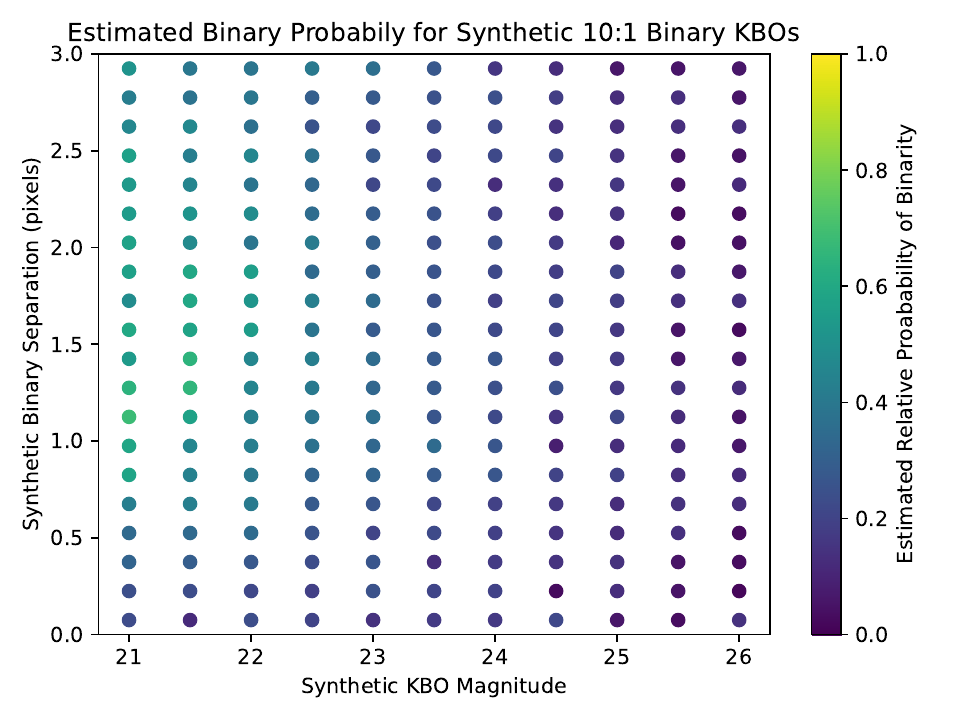}}
\caption{
    Recovery efficiency of simulated binary KBOs with SSOLS-style synthetic data, assuming jitter of 0.1 pixels and F606W-F814W color of 0.72.
    Each panel corresponds to 1000 simulations with a different primary/secondary flux ratio.
    The synthetic data was processed in the same way as the actual SSOLS data.
    The separations of the synthetic binaries were varied uniformly from 0 to 3 pixels, but binned and averaged here for clarity.
    Recovery efficiency was as expected better for brighter satellites farther away, 
    with very few (but still some) 10:1 flux ratio binaries detectable.
    None of the synthetic single-PSF cases were detected as a false-positive binary at 95\% confidence.
\label{fig:ssolssynth}
}
\end{figure*}

Our simulation results can be seen in Figure \ref{fig:ssolssynth}.
Equal-brightness binaries (top left of Figure \ref{fig:ssolssynth}) with V=21 combined brightness 
(roughly 400 km diameter for both objects at 40 AU and 5\% albedo, same for following estimates)
can be easily detected at separations as close as a half WFC3 pixel (roughly 600 km at 40 AU).
This drops off with brightness, with a full pixel of separation being needed at V=22.5 
($\approx$200 km diameters, $\approx$1200 km separation),
and effectively no equal-brightness blended binaries were detectable at a combined magnitude less than V=24.5
(roughly 80 km diameter for both objects).
HST/WFC3 is thus able to probe to the Roche limit for equal-brightness binaries larger than roughly 400 km,
but is ineffective at detecting equal-brightness binaries smaller than 80 km diameter at separations less
than at least 90 radii apart.
This is crucial to understanding the missing tight binary population, as 
(523764) 2014 WC$_{510}$ was observed by stellar occultation to have a separation of 349 km, $<$4 radii apart \citep{2020PSJ.....1...48L},
and 2011 JY$_{31}$ was observed by New Horizons to have a separation of 198 km, a few radii apart \citep{2022PSJ.....3...46W}.
In addition, the simulations of \citet{2012Icar..220..947P} showed there should be a large number of binary KBOs at separations of
a few radii, as that is where tidal forces circularized orbits and stopped semimajor axis decay.
HST/WFC3 is thus generally insensitive to a potentially large population of binary KBOs that is both known to 
exist from other detection methods, and predicted by theory.

Binary KBOs with a 2:1 brightness ratio fared only slightly worse, with a similar binary recovery rate for V=21
(primary roughly 500 km diameter, secondary roughly 300 km), down to half a pixel (roughly 600 km),
and V=22.5 (primary roughly 240 km, secondary 170 km) also needing a full pixel of separation (1200 km).
However, the drop to a 2:1 flux ratio increased the threshold of any blended binary detection at 3 pixels 
from V=24.5 to V=24.0 (120 and 85 km diameters).
At 5:1 brightness ratio, the minimum combined brightness for the detection of a satellite was V=23
(primary diameter about 210 km, secondary 95 km).
At 10:1 brightness ratio, the minimum combined brightness for the detection of a satellite was V=21.5
(primary diameter about 440 km, secondary 140 km).

\section{Results and Discussion}

We have performed initial binary detection and for all the data in program GO-15648.
The results of our analysis of binary probability are listed in Table \ref{tab:list}, and compared to the known
binary KBOs in Figure \ref{fig:cudist}.
Of the 198 objects that were successfully observed, we identified 23 as being visually separated binaries,
a further 19 as blended binaries with a probability of being binary of being $>$95\% (per the formulation above),
and 156 that appeared as singular objects to our analysis.
Our blended binary analysis was thus able to almost double the yield of binary KBOs in the SSOLS dataset.

Our detection of a large number of blended-PSF binary KBOs implies that binary KBOs at apparent separations 
at or just below the detection limit of HST/WFC3 may be very common.
This argument is further strengthened by our synthetic KBO analysis which shows that close binary KBOs dimmer
than V=24 and with flux ratios larger than 5:1 are very hard for WFC3 to detect, even with apparent separations
greater than one pixel.
A full census of the binary fraction in the Kuiper Belt is thus very difficult from direct imaging alone.
Ground-based adaptive optics systems are even more flux limited than HST \citep{2011Icar..213..678G} 
and cannot probe binaries that are fainter or closer than those detectable with HST/WFC3.
The James Webb Space Telescope (JWST) with the NIRCam instrument only offers a marginal increase in angular resolution
over WFC3 ($\approx$30 milliarcseconds/pixel versus $\approx$40 milliarcseconds/pixel for WFC3),
but may be able to detect fainter KBO satellites due to its larger collecting area.

The strongest evidence that very tight binary KBOs may be common comes from two unconventional binary detection
methods, a stellar occultation and direct spacecraft observations.
The 3:2 Neptune resonant KBO
(523764) 2014 WC$_{510}$ was observed by stellar occultation to have a separation of 349 km \citep{2020PSJ.....1...48L},
which corresponds to an apparent separation of 13 milliarcseconds, or less than a 1/3 of a WFC3 pixel.
The New Horizons spacecraft observed five KBOs with the LORRI camera at a distance of less than 0.1 astronomical units,
and two of them appeared to be very close binaries, 2011 JY$_{31}$ and 2014 OS$_{393}$ \citep{2022PSJ.....3...46W}.
2011 JY$_{31}$ was resolved as a blended binary in both LORRI epochs, and the combination of that with the
assumption of the orbital period being equal to the lightcurve period of 46 hours allowed the mutual orbit separation
to be fit as 199 km \citep{2022PSJ.....3...46W}.
2014 OS$_{393}$ was only observed to be binary in one of the two epochs, and \citet{2022PSJ.....3...46W} estimated
its apparent separation to be $\approx$150 km.
These separations would be $<$0.2 WFC3 pixels as seen by HST, and sure enough, neither 2011 JY$_{31}$ nor
2014 OS$_{393}$ appears obviously binary in HST WFC3 imaging \citep{2022PSJ.....3...46W}.
While these unconventional detections are hard to reproduce, they do strongly argue that binary KBOs with separations
below the detection threshold of WFC3 should be common.
These tight binaries may be a result of direct formation from the protoplanetary disk by the streaming instability 
\citep[e.g.][]{2017NatAs...1E..88F,2019NatAs...3..808N},
or the result of post-formation orbital evolution \citep[e.g.][]{2012Icar..220..947P},
or some combination of those processes.
The discovery of both of these very tight binary KBOs ($<$400 km) with unconventional methods,
and the large number of 1000-3000 km binaries that we found in the SSOLS dataset,
shows that there is likely to be a substatial number of Kuiper Belt binaries with
300-1000 km separations.

The inclinations of binary KBOs to their orbits has been shown to be very preferentially prograde \citep{2019Icar..334...62G}.
This provides an important constraint on the formation of binary KBOs, as it strongly supports the formation
of KBO binaries by SI \citep{2019NatAs...3..808N}.
However, this test has only been performed for relatively wide KBO binaries \citep{2019Icar..334...62G},
as very few tight binaries have fully determined orbits.
Formation models \citep[e.g.][]{2019NatAs...3..808N} show that most binary KBOs formed wider and more eccentric than
the observed binary KBOs \citep{2019Icar..334...62G}, before the mutual orbits evolved under perturbations
to tighter, more circular orbits \citep{2012Icar..220..947P}.
Models imply that the prograde preference should be preserved under that mutual orbit
evolution \citep{2012Icar..220..947P}, but that has not yet been proven, and can only be tested by measuring
the orbits of tight binary KBOs.
Future observations of the blended binary KBOs discovered in this program could fit their mutual orbits
which would be very useful to test if this prograde preference is indeed preserved to much tighter separations.

\section{Future Observations and Analysis}

HST/WFC3 has historically been the high-resolution instrument available for this work \citep[e.g.][and citations therein]{2019Icar..334...62G}, 
but JWST/NIRCam does offer slightly higher resolution than WFC3 \citep{2023PASP..135b8001R}, with a much larger collecting area.
However, JWST observations of KBOs is challenging, as the design of the spacecraft prevents it from observing the
Kuiper Belt except at the quadratures, and after the micrometorite constraint imposed in Cycle 2,
only the post-opposition quadrature window is available.
Any KBO must therefore be observed by JWST within a narrow, roughly two week window, away from the opposition surge that 
pushes faint KBOs to their maximum brightness, and with slow moving background stars that may be confused for KBO satellites.
In addition, the narrow observing window of quadrature-only space telescopes, like JWST and the Roman Space Telescope,
makes recover of the mutual orbits of binary KBOs effectively impossible, as they cannot cover the required temporal range
\citep{2008Icar..197..260G}.
The Keck laser adaptive optics system has been used extensively for recovery of binary KBOs at slightly lower resolution than HST \citep{2011Icar..213..678G}.
Future ground-based facilities like the Extremely Large Telescope (ELT) and the Thirty Meter Telescope (TMT)
may be able to directly resolve even tighter binaries, with their larger collecting areas being able to compensate for the
throughput losses inherent in adaptive optics systems.
In addition, both of those future large telescopes are planned to use multi-point laser adaptive optics systems \citep{2016SPIE.9909E..2DD,2014SPIE.9148E..0XB}
which would enable them to observe KBOs without the need of a stellar appulse for tip/tilt correction (as used by Keck), 
increasing their operational flexibility.

The most promising pathway to surveying the very close binary population of the Kuiper Belt may be through stellar occultations.
As \citet{2020PSJ.....1...48L} has shown, stellar occultation can detect binary KBOs down to arbitrary separations,
and the use of the Gaia DR3 star catalog \citep{2021A&A...649A...1G} allows these occultations to be targeted with great precision \citep{2018AJ....156...20P}.
The advent of low cost cameras with GPS timing precision enables stellar occultations to be observed with small, highly mobile
telescopes, greatly increasing the coverage possible for a given occultation event.
Given that 2/6 KBOs observed up close by New Horizons (including Arrokoth) were apparently binaries with less than 200 km separation,
it would not take many stellar occultations to find additional very close binary KBOs.

For future analysis of this and similar datasets, we plan to enhance binary detection with the use of Machine Learning (ML). 
The current results are based on shifting Tiny-Tim PSFs and assumptions about the noise properties of WFC3 images. 
To use ML, we would create a large amount of synthetic single and binary KBOs (much larger than used for our calibration here)
and then train a neural network to distinguish between them.
This could result in a much more effective detection of barely-resolvable binary KBOs, and would enable the application of this
method to historical observations of KBOs by HST to find even more blended-PSF binaries.

\section{Conclusions}

We performed a PSF-fitting process on 198 KBOs observed with HST/WFC3 and found that while 11.6\% appeared to be visually
separated binaries, a further 9.6\% were blended-PSF binaries.
Our results show that binary KBOs at or below the detection threshold of HST appear to be at least as common as binary KBOs
that appear to be well separated in WFC3/UVIS images.
This was confirmed by our application of the same fitting method to synthetic data, which found that binary KBOs fainter than
V=24 or with primary/secondary flux ratios larger than 5:1 are very unlikely to be detectable with HST.
These results appear to support the idea that the Kuiper Belt was formed through the Streaming Instability process,
and that most if not all KBOs were born as binary systems.

\begin{acknowledgments}
This work was supported by HST program GO-15648. 
HST data was obtained from the Space Telescope Science Institute, 
which is operated by the Association of Universities for Research in Astronomy, Inc., under NASA contract NAS 5–26555.
\end{acknowledgments}

\software{
    Astropy \citep{2022ApJ...935..167A},
    Scipy \citep{2020NatMe..17..261V},
    Photutils \citep{2022zndo...6825092B},
    Spiceypy \citep{2021zndo...4883901A},
    Tiny Tim \citep{1995ASPC...77..349K},
}

\bibliography{refs}{}

\begin{thebibliography}{}
\expandafter\ifx\csname natexlab\endcsname\relax\def\natexlab#1{#1}\fi
\providecommand{\url}[1]{\href{#1}{#1}}
\providecommand{\dodoi}[1]{doi:~\href{http://doi.org/#1}{\nolinkurl{#1}}}
\providecommand{\doeprint}[1]{\href{http://ascl.net/#1}{\nolinkurl{http://ascl.net/#1}}}
\providecommand{\doarXiv}[1]{\href{https://arxiv.org/abs/#1}{\nolinkurl{https://arxiv.org/abs/#1}}}

\bibitem[{{Annex} {et~al.}(2021){Annex}, {Pearson}, {Seignovert}, {Carcich},
  {Eichhorn}, {Mapel}, {Von Forstner}, {McAuliffe}, {Diaz Del Rio}, {Berry},
  {Aye}, {Stefko}, {De Val-Borro}, {Kulumani}, {Murakami}, {Niemeyer},
  {Medley}, \& {Margot}}]{2021zndo...4883901A}
{Annex}, A., {Pearson}, B., {Seignovert}, B., {et~al.} 2021,
  {AndrewAnnex/SpiceyPy: SpiceyPy 4.0.1}, v4.0.1, Zenodo,  Zenodo,
  \dodoi{10.5281/zenodo.4883901}

\bibitem[{{Astropy Collaboration} {et~al.}(2022){Astropy Collaboration},
  {Price-Whelan}, {Lim}, {Earl}, {Starkman}, {Bradley}, {Shupe}, {Patil},
  {Corrales}, {Brasseur}, {N{\"o}the}, {Donath}, {Tollerud}, {Morris},
  {Ginsburg}, {Vaher}, {Weaver}, {Tocknell}, {Jamieson}, {van Kerkwijk},
  {Robitaille}, {Merry}, {Bachetti}, {G{\"u}nther}, {Aldcroft},
  {Alvarado-Montes}, {Archibald}, {B{\'o}di}, {Bapat}, {Barentsen},
  {Baz{\'a}n}, {Biswas}, {Boquien}, {Burke}, {Cara}, {Cara}, {Conroy},
  {Conseil}, {Craig}, {Cross}, {Cruz}, {D'Eugenio}, {Dencheva}, {Devillepoix},
  {Dietrich}, {Eigenbrot}, {Erben}, {Ferreira}, {Foreman-Mackey}, {Fox},
  {Freij}, {Garg}, {Geda}, {Glattly}, {Gondhalekar}, {Gordon}, {Grant},
  {Greenfield}, {Groener}, {Guest}, {Gurovich}, {Handberg}, {Hart},
  {Hatfield-Dodds}, {Homeier}, {Hosseinzadeh}, {Jenness}, {Jones}, {Joseph},
  {Kalmbach}, {Karamehmetoglu}, {Ka{\l}uszy{\'n}ski}, {Kelley}, {Kern},
  {Kerzendorf}, {Koch}, {Kulumani}, {Lee}, {Ly}, {Ma}, {MacBride}, {Maljaars},
  {Muna}, {Murphy}, {Norman}, {O'Steen}, {Oman}, {Pacifici}, {Pascual},
  {Pascual-Granado}, {Patil}, {Perren}, {Pickering}, {Rastogi}, {Roulston},
  {Ryan}, {Rykoff}, {Sabater}, {Sakurikar}, {Salgado}, {Sanghi}, {Saunders},
  {Savchenko}, {Schwardt}, {Seifert-Eckert}, {Shih}, {Jain}, {Shukla}, {Sick},
  {Simpson}, {Singanamalla}, {Singer}, {Singhal}, {Sinha}, {Sip{\H{o}}cz},
  {Spitler}, {Stansby}, {Streicher}, {{\v{S}}umak}, {Swinbank}, {Taranu},
  {Tewary}, {Tremblay}, {de Val-Borro}, {Van Kooten}, {Vasovi{\'c}}, {Verma},
  {de Miranda Cardoso}, {Williams}, {Wilson}, {Winkel}, {Wood-Vasey}, {Xue},
  {Yoachim}, {Zhang}, {Zonca}, \& {Astropy Project
  Contributors}}]{2022ApJ...935..167A}
{Astropy Collaboration}, {Price-Whelan}, A.~M., {Lim}, P.~L., {et~al.} 2022,
  \apj, 935, 167, \dodoi{10.3847/1538-4357/ac7c74}

\bibitem[{{Bannister} {et~al.}(2018){Bannister}, {Gladman}, {Kavelaars},
  {Petit}, {Volk}, {Chen}, {Alexandersen}, {Gwyn}, {Schwamb}, {Ashton},
  {Benecchi}, {Cabral}, {Dawson}, {Delsanti}, {Fraser}, {Granvik},
  {Greenstreet}, {Guilbert-Lepoutre}, {Ip}, {Jakubik}, {Jones}, {Kaib},
  {Lacerda}, {Van Laerhoven}, {Lawler}, {Lehner}, {Lin}, {Lykawka}, {Marsset},
  {Murray-Clay}, {Pike}, {Rousselot}, {Shankman}, {Thirouin}, {Vernazza}, \&
  {Wang}}]{2018ApJS..236...18B}
{Bannister}, M.~T., {Gladman}, B.~J., {Kavelaars}, J.~J., {et~al.} 2018, \apjs,
  236, 18, \dodoi{10.3847/1538-4365/aab77a}

\bibitem[{{Benecchi} {et~al.}(2019){Benecchi}, {Borncamp}, {Parker}, {Buie},
  {Noll}, {Binzel}, {Stern}, {Verbiscer}, {Kavelaars}, {Zangari}, {Spencer}, \&
  {Weaver}}]{2019Icar..334...22B}
{Benecchi}, S.~D., {Borncamp}, D., {Parker}, A.~H., {et~al.} 2019, \icarus,
  334, 22, \dodoi{10.1016/j.icarus.2019.01.025}

\bibitem[{{Boyer} {et~al.}(2014){Boyer}, {Adkins}, {Andersen}, {Atwood}, {Bo},
  {Byrnes}, {Caputa}, {Cavaco}, {Ellerbroek}, {Gilles}, {Gregory}, {Herriot},
  {Hickson}, {Ljusic}, {Manter}, {Marois}, {Ot{\'a}rola}, {Pag{\`e}s},
  {Schoeck}, {Sinquin}, {Smith}, {Spano}, {Szeto}, {Tang}, {Travouillon},
  {V{\'e}ran}, {Wang}, \& {Wei}}]{2014SPIE.9148E..0XB}
{Boyer}, C., {Adkins}, S., {Andersen}, D.~R., {et~al.} 2014, in Society of
  Photo-Optical Instrumentation Engineers (SPIE) Conference Series, Vol. 9148,
  Adaptive Optics Systems IV, ed. E.~{Marchetti}, L.~M. {Close}, \& J.-P.
  {Vran}, 91480X, \dodoi{10.1117/12.2056863}

\bibitem[{{Bradley} {et~al.}(2022){Bradley}, {Sip{\H{o}}cz}, {Robitaille},
  {Tollerud}, {Vin{\'\i}cius}, {Deil}, {Barbary}, {Wilson}, {Busko}, {Donath},
  {G{\"u}nther}, {Cara}, {Lim}, {Me{\ss}linger}, {Conseil}, {Bostroem},
  {Droettboom}, {Bray}, {Andersen Bratholm}, {Barentsen}, {Craig}, {Rathi},
  {Pascual}, {Perren}, {Georgiev}, {De Val-Borro}, {Kerzendorf}, {Bach},
  {Quint}, \& {Souchereau}}]{2022zndo...6825092B}
{Bradley}, L., {Sip{\H{o}}cz}, B., {Robitaille}, T., {et~al.} 2022,
  {astropy/photutils: 1.5.0}, 1.5.0, Zenodo,  Zenodo,
  \dodoi{10.5281/zenodo.6825092}

\bibitem[{{Canup}(2011)}]{2011AJ....141...35C}
{Canup}, R.~M. 2011, \aj, 141, 35, \dodoi{10.1088/0004-6256/141/2/35}

\bibitem[{{Diolaiti} {et~al.}(2016){Diolaiti}, {Ciliegi}, {Abicca}, {Agapito},
  {Arcidiacono}, {Baruffolo}, {Bellazzini}, {Biliotti}, {Bonaglia}, {Bregoli},
  {Briguglio}, {Brissaud}, {Busoni}, {Carbonaro}, {Carlotti}, {Cascone},
  {Correia}, {Cortecchia}, {Cosentino}, {De Caprio}, {de Pascale}, {De Rosa},
  {Del Vecchio}, {Delboulb{\'e}}, {Di Rico}, {Esposito}, {Fantinel},
  {Feautrier}, {Felini}, {Ferruzzi}, {Fini}, {Fiorentino}, {Foppiani}, {Ghigo},
  {Giordano}, {Giro}, {Gluck}, {H{\'e}nault}, {Jocou}, {Kerber}, {La Penna},
  {Lafrasse}, {Lauria}, {le Coarer}, {Le Louarn}, {Lombini}, {Magnard},
  {Maiorano}, {Mannucci}, {Mapelli}, {Marchetti}, {Maurel}, {Michaud},
  {Morgante}, {Moulin}, {Oberti}, {Pareschi}, {Patti}, {Puglisi}, {Rabou},
  {Ragazzoni}, {Ramsay}, {Riccardi}, {Ricciardi}, {Riva}, {Rochat}, {Roussel},
  {Roux}, {Salasnich}, {Saracco}, {Schreiber}, {Spavone}, {Stadler}, {Sztefek},
  {Ventura}, {V{\'e}rinaud}, {Xompero}, {Fontana}, \&
  {Zerbi}}]{2016SPIE.9909E..2DD}
{Diolaiti}, E., {Ciliegi}, P., {Abicca}, R., {et~al.} 2016, in Society of
  Photo-Optical Instrumentation Engineers (SPIE) Conference Series, Vol. 9909,
  Adaptive Optics Systems V, ed. E.~{Marchetti}, L.~M. {Close}, \& J.-P.
  {V{\'e}ran}, 99092D, \dodoi{10.1117/12.2234585}

\bibitem[{{Elliot} {et~al.}(2005){Elliot}, {Kern}, {Clancy}, {Gulbis},
  {Millis}, {Buie}, {Wasserman}, {Chiang}, {Jordan}, {Trilling}, \&
  {Meech}}]{2005AJ....129.1117E}
{Elliot}, J.~L., {Kern}, S.~D., {Clancy}, K.~B., {et~al.} 2005, \aj, 129, 1117,
  \dodoi{10.1086/427395}

\bibitem[{{Fraser} {et~al.}(2017){Fraser}, {Bannister}, {Pike}, {Marsset},
  {Schwamb}, {Kavelaars}, {Lacerda}, {Nesvorn{\'y}}, {Volk}, {Delsanti},
  {Benecchi}, {Lehner}, {Noll}, {Gladman}, {Petit}, {Gwyn}, {Chen}, {Wang},
  {Alexandersen}, {Burdullis}, {Sheppard}, \& {Trujillo}}]{2017NatAs...1E..88F}
{Fraser}, W.~C., {Bannister}, M.~T., {Pike}, R.~E., {et~al.} 2017, Nature
  Astronomy, 1, 0088, \dodoi{10.1038/s41550-017-0088}

\bibitem[{{Gaia Collaboration} {et~al.}(2021){Gaia Collaboration}, {Brown},
  {Vallenari}, {Prusti}, {de Bruijne}, {Babusiaux}, {Biermann}, {Creevey},
  {Evans}, {Eyer}, \& et~al.}]{2021A&A...649A...1G}
{Gaia Collaboration}, {Brown}, A.~G.~A., {Vallenari}, A., {et~al.} 2021, \aap,
  649, A1, \dodoi{10.1051/0004-6361/202039657}

\bibitem[{{Grundy} {et~al.}(2008){Grundy}, {Noll}, {Virtanen}, {Muinonen},
  {Kern}, {Stephens}, {Stansberry}, {Levison}, \&
  {Spencer}}]{2008Icar..197..260G}
{Grundy}, W.~M., {Noll}, K.~S., {Virtanen}, J., {et~al.} 2008, \icarus, 197,
  260, \dodoi{10.1016/j.icarus.2008.04.004}

\bibitem[{{Grundy} {et~al.}(2011){Grundy}, {Noll}, {Nimmo}, {Roe}, {Buie},
  {Porter}, {Benecchi}, {Stephens}, {Levison}, \&
  {Stansberry}}]{2011Icar..213..678G}
{Grundy}, W.~M., {Noll}, K.~S., {Nimmo}, F., {et~al.} 2011, \icarus, 213, 678,
  \dodoi{10.1016/j.icarus.2011.03.012}

\bibitem[{{Grundy} {et~al.}(2019){Grundy}, {Noll}, {Roe}, {Buie}, {Porter},
  {Parker}, {Nesvorn{\'y}}, {Levison}, {Benecchi}, {Stephens}, \&
  {Trujillo}}]{2019Icar..334...62G}
{Grundy}, W.~M., {Noll}, K.~S., {Roe}, H.~G., {et~al.} 2019, \icarus, 334, 62,
  \dodoi{10.1016/j.icarus.2019.03.035}

\bibitem[{{Grundy} {et~al.}(2020){Grundy}, {Bird}, {Britt}, {Cook},
  {Cruikshank}, {Howett}, {Krijt}, {Linscott}, {Olkin}, {Parker}, {Protopapa},
  {Ruaud}, {Umurhan}, {Young}, {Dalle Ore}, {Kavelaars}, {Keane}, {Pendleton},
  {Porter}, {Scipioni}, {Spencer}, {Stern}, {Verbiscer}, {Weaver}, {Binzel},
  {Buie}, {Buratti}, {Cheng}, {Earle}, {Elliott}, {Gabasova}, {Gladstone},
  {Hill}, {Horanyi}, {Jennings}, {Lunsford}, {McComas}, {McKinnon}, {McNutt},
  {Moore}, {Parker}, {Quirico}, {Reuter}, {Schenk}, {Schmitt}, {Showalter},
  {Singer}, {Weigle}, \& {Zangari}}]{2020Sci...367.3705G}
{Grundy}, W.~M., {Bird}, M.~K., {Britt}, D.~T., {et~al.} 2020, Science, 367,
  aay3705, \dodoi{10.1126/science.aay3705}

\bibitem[{{Krist}(1995)}]{1995ASPC...77..349K}
{Krist}, J. 1995, in Astronomical Society of the Pacific Conference Series,
  Vol.~77, Astronomical Data Analysis Software and Systems IV, ed. R.~A.
  {Shaw}, H.~E. {Payne}, \& J.~J.~E. {Hayes}, 349

\bibitem[{{Leiva} {et~al.}(2020){Leiva}, {Buie}, {Keller}, {Wasserman},
  {Kavelaars}, {Bridges}, {Haley}, {Strauss}, {Wilde}, {Weryk}, {Kervella},
  {Baker}, {Bock}, {Conway}, {Cota}, {Estes}, {Garc{\'\i}a}, {Kehrli},
  {McCandless}, {McCandless}, {Self}, {Settlemire}, {Swanson}, {Thompson}, \&
  {Wise}}]{2020PSJ.....1...48L}
{Leiva}, R., {Buie}, M.~W., {Keller}, J.~M., {et~al.} 2020, \psj, 1, 48,
  \dodoi{10.3847/PSJ/abb23d}

\bibitem[{{Mao} {et~al.}(2021){Mao}, {McKinnon}, {Singer}, {Keane}, {Beyer},
  {Greenstreet}, {Robbins}, {Schenk}, {Moore}, {Stern}, {Weaver}, {Spencer}, \&
  {Olkin}}]{2021JGRE..12606961M}
{Mao}, X., {McKinnon}, W.~B., {Singer}, K.~N., {et~al.} 2021, Journal of
  Geophysical Research (Planets), 126, e06961, \dodoi{10.1029/2021JE006961}

\bibitem[{{McKinnon} {et~al.}(2020){McKinnon}, {Richardson}, {Marohnic},
  {Keane}, {Grundy}, {Hamilton}, {Nesvorn{\'y}}, {Umurhan}, {Lauer}, {Singer},
  {Stern}, {Weaver}, {Spencer}, {Buie}, {Moore}, {Kavelaars}, {Lisse}, {Mao},
  {Parker}, {Porter}, {Showalter}, {Olkin}, {Cruikshank}, {Elliott},
  {Gladstone}, {Parker}, {Verbiscer}, {Young}, \& {New Horizons Science
  Team}}]{2020Sci...367.6620M}
{McKinnon}, W.~B., {Richardson}, D.~C., {Marohnic}, J.~C., {et~al.} 2020,
  Science, 367, aay6620, \dodoi{10.1126/science.aay6620}

\bibitem[{{Nesvorn{\'y}} {et~al.}(2019){Nesvorn{\'y}}, {Li}, {Youdin}, {Simon},
  \& {Grundy}}]{2019NatAs...3..808N}
{Nesvorn{\'y}}, D., {Li}, R., {Youdin}, A.~N., {Simon}, J.~B., \& {Grundy},
  W.~M. 2019, Nature Astronomy, 3, 808, \dodoi{10.1038/s41550-019-0806-z}

\bibitem[{{Petit} {et~al.}(2011){Petit}, {Kavelaars}, {Gladman}, {Jones},
  {Parker}, {Van Laerhoven}, {Nicholson}, {Mars}, {Rousselot}, {Mousis},
  {Marsden}, {Bieryla}, {Taylor}, {Ashby}, {Benavidez}, {Campo Bagatin}, \&
  {Bernabeu}}]{2011AJ....142..131P}
{Petit}, J.~M., {Kavelaars}, J.~J., {Gladman}, B.~J., {et~al.} 2011, \aj, 142,
  131, \dodoi{10.1088/0004-6256/142/4/131}

\bibitem[{{Porter} \& {Grundy}(2012)}]{2012Icar..220..947P}
{Porter}, S.~B., \& {Grundy}, W.~M. 2012, \icarus, 220, 947,
  \dodoi{10.1016/j.icarus.2012.06.034}

\bibitem[{{Porter} {et~al.}(2018){Porter}, {Buie}, {Parker}, {Spencer},
  {Benecchi}, {Tanga}, {Verbiscer}, {Kavelaars}, {Gwyn}, {Young}, {Weaver},
  {Olkin}, {Parker}, \& {Stern}}]{2018AJ....156...20P}
{Porter}, S.~B., {Buie}, M.~W., {Parker}, A.~H., {et~al.} 2018, \aj, 156, 20,
  \dodoi{10.3847/1538-3881/aac2e1}

\bibitem[{{Rieke} {et~al.}(2023){Rieke}, {Kelly}, {Misselt}, {Stansberry},
  {Boyer}, {Beatty}, {Egami}, {Florian}, {Greene}, {Hainline}, {Leisenring},
  {Roellig}, {Schlawin}, {Sun}, {Tinnin}, {Williams}, {Willmer}, {Wilson},
  {Clark}, {Rohrbach}, {Brooks}, {Canipe}, {Correnti}, {DiFelice}, {Gennaro},
  {Girard}, {Hartig}, {Hilbert}, {Koekemoer}, {Nikolov}, {Pirzkal}, {Rest},
  {Robberto}, {Sunnquist}, {Telfer}, {Wu}, {Ferry}, {Lewis}, {Baum},
  {Beichman}, {Doyon}, {Dressler}, {Eisenstein}, {Ferrarese}, {Hodapp},
  {Horner}, {Jaffe}, {Johnstone}, {Krist}, {Martin}, {McCarthy}, {Meyer},
  {Rieke}, {Trauger}, \& {Young}}]{2023PASP..135b8001R}
{Rieke}, M.~J., {Kelly}, D.~M., {Misselt}, K., {et~al.} 2023, \pasp, 135,
  028001, \dodoi{10.1088/1538-3873/acac53}

\bibitem[{{Robbins} {et~al.}(2017){Robbins}, {Singer}, {Bray}, {Schenk},
  {Lauer}, {Weaver}, {Runyon}, {McKinnon}, {Beyer}, {Porter}, {White},
  {Hofgartner}, {Zangari}, {Moore}, {Young}, {Spencer}, {Binzel}, {Buie},
  {Buratti}, {Cheng}, {Grundy}, {Linscott}, {Reitsema}, {Reuter}, {Showalter},
  {Tyler}, {Olkin}, {Ennico}, {Stern}, \& {New Horizons
  Lorri}}]{2017Icar..287..187R}
{Robbins}, S.~J., {Singer}, K.~N., {Bray}, V.~J., {et~al.} 2017, \icarus, 287,
  187, \dodoi{10.1016/j.icarus.2016.09.027}

\bibitem[{{Showalter} {et~al.}(2021){Showalter}, {Benecchi}, {Buie}, {Grundy},
  {Keane}, {Lisse}, {Olkin}, {Porter}, {Robbins}, {Singer}, {Verbiscer},
  {Weaver}, {Zangari}, {Hamilton}, {Kaufmann}, {Lauer}, {Mehoke}, {Mehoke},
  {Spencer}, {Throop}, {Parker}, {Stern}, {New Horizons Geology}, \&
  Team}]{2021Icar..35614098S}
{Showalter}, M.~R., {Benecchi}, S.~D., {Buie}, M.~W., {et~al.} 2021, \icarus,
  356, 114098, \dodoi{10.1016/j.icarus.2020.114098}

\bibitem[{{Spencer} {et~al.}(2020){Spencer}, {Stern}, {Moore}, {Weaver},
  {Singer}, {Olkin}, {Verbiscer}, {McKinnon}, {Parker}, {Beyer}, {Keane},
  {Lauer}, {Porter}, {White}, {Buratti}, {El-Maarry}, {Lisse}, {Parker},
  {Throop}, {Robbins}, {Umurhan}, {Binzel}, {Britt}, {Buie}, {Cheng},
  {Cruikshank}, {Elliott}, {Gladstone}, {Grundy}, {Hill}, {Horanyi},
  {Jennings}, {Kavelaars}, {Linscott}, {McComas}, {McNutt}, {Protopapa},
  {Reuter}, {Schenk}, {Showalter}, {Young}, {Zangari}, {Abedin},
  {Beddingfield}, {Benecchi}, {Bernardoni}, {Bierson}, {Borncamp}, {Bray},
  {Chaikin}, {Dhingra}, {Fuentes}, {Fuse}, {Gay}, {Gwyn}, {Hamilton},
  {Hofgartner}, {Holman}, {Howard}, {Howett}, {Karoji}, {Kaufmann}, {Kinczyk},
  {May}, {Mountain}, {P{\"a}tzold}, {Petit}, {Piquette}, {Reid}, {Reitsema},
  {Runyon}, {Sheppard}, {Stansberry}, {Stryk}, {Tanga}, {Tholen}, {Trilling},
  \& {Wasserman}}]{2020Sci...367.3999S}
{Spencer}, J.~R., {Stern}, S.~A., {Moore}, J.~M., {et~al.} 2020, Science, 367,
  aay3999, \dodoi{10.1126/science.aay3999}

\bibitem[{{Stern} {et~al.}(2015){Stern}, {Bagenal}, {Ennico}, {Gladstone},
  {Grundy}, {McKinnon}, {Moore}, {Olkin}, {Spencer}, {Weaver}, {Young},
  {Andert}, {Andrews}, {Banks}, {Bauer}, {Bauman}, {Barnouin}, {Bedini},
  {Beisser}, {Beyer}, {Bhaskaran}, {Binzel}, {Birath}, {Bird}, {Bogan},
  {Bowman}, {Bray}, {Brozovic}, {Bryan}, {Buckley}, {Buie}, {Buratti},
  {Bushman}, {Calloway}, {Carcich}, {Cheng}, {Conard}, {Conrad}, {Cook},
  {Cruikshank}, {Custodio}, {Dalle Ore}, {Deboy}, {Dischner}, {Dumont},
  {Earle}, {Elliott}, {Ercol}, {Ernst}, {Finley}, {Flanigan}, {Fountain},
  {Freeze}, {Greathouse}, {Green}, {Guo}, {Hahn}, {Hamilton}, {Hamilton},
  {Hanley}, {Harch}, {Hart}, {Hersman}, {Hill}, {Hill}, {Hinson}, {Holdridge},
  {Horanyi}, {Howard}, {Howett}, {Jackman}, {Jacobson}, {Jennings}, {Kammer},
  {Kang}, {Kaufmann}, {Kollmann}, {Krimigis}, {Kusnierkiewicz}, {Lauer}, {Lee},
  {Lindstrom}, {Linscott}, {Lisse}, {Lunsford}, {Mallder}, {Martin}, {McComas},
  {McNutt}, {Mehoke}, {Mehoke}, {Melin}, {Mutchler}, {Nelson}, {Nimmo},
  {Nunez}, {Ocampo}, {Owen}, {Paetzold}, {Page}, {Parker}, {Parker},
  {Pelletier}, {Peterson}, {Pinkine}, {Piquette}, {Porter}, {Protopapa},
  {Redfern}, {Reitsema}, {Reuter}, {Roberts}, {Robbins}, {Rogers}, {Rose},
  {Runyon}, {Retherford}, {Ryschkewitsch}, {Schenk}, {Schindhelm}, {Sepan},
  {Showalter}, {Singer}, {Soluri}, {Stanbridge}, {Steffl}, {Strobel}, {Stryk},
  {Summers}, {Szalay}, {Tapley}, {Taylor}, {Taylor}, {Throop}, {Tsang},
  {Tyler}, {Umurhan}, {Verbiscer}, {Versteeg}, {Vincent}, {Webbert}, {Weidner},
  {Weigle}, {White}, {Whittenburg}, {Williams}, {Williams}, {Williams},
  {Woods}, {Zangari}, \& {Zirnstein}}]{2015Sci...350.1815S}
{Stern}, S.~A., {Bagenal}, F., {Ennico}, K., {et~al.} 2015, Science, 350,
  aad1815, \dodoi{10.1126/science.aad1815}

\bibitem[{{Thirouin} \& {Sheppard}(2018)}]{2018AJ....155..248T}
{Thirouin}, A., \& {Sheppard}, S.~S. 2018, \aj, 155, 248,
  \dodoi{10.3847/1538-3881/aac0ff}

\bibitem[{{Thirouin} \& {Sheppard}(2019)}]{2019AJ....157..228T}
---. 2019, \aj, 157, 228, \dodoi{10.3847/1538-3881/ab18a9}

\bibitem[{{Virtanen} {et~al.}(2020){Virtanen}, {Gommers}, {Oliphant},
  {Haberland}, {Reddy}, {Cournapeau}, {Burovski}, {Peterson}, {Weckesser},
  {Bright}, {van der Walt}, {Brett}, {Wilson}, {Millman}, {Mayorov}, {Nelson},
  {Jones}, {Kern}, {Larson}, {Carey}, {Polat}, {Feng}, {Moore}, {VanderPlas},
  {Laxalde}, {Perktold}, {Cimrman}, {Henriksen}, {Quintero}, {Harris},
  {Archibald}, {Ribeiro}, {Pedregosa}, {van Mulbregt}, \& {SciPy 1. 0
  Contributors}}]{2020NatMe..17..261V}
{Virtanen}, P., {Gommers}, R., {Oliphant}, T.~E., {et~al.} 2020, Nature
  Methods, 17, 261, \dodoi{10.1038/s41592-019-0686-2}

\bibitem[{{Weaver} {et~al.}(2022){Weaver}, {Porter}, {Spencer}, \& {The New
  Horizons Science Team}}]{2022PSJ.....3...46W}
{Weaver}, H.~A., {Porter}, S.~B., {Spencer}, J.~R., \& {The New Horizons
  Science Team}. 2022, \psj, 3, 46, \dodoi{10.3847/PSJ/ac4cb7}

\end{thebibliography}
\bibliographystyle{aasjournal}

\end{document}